# Metallicity and age effects on lithium depletion in solar analogues

Giulia Martos,★ Jorge Meléndez, Anne Rathsam and Gabriela Carvalho Silva

*Departamento de Astronomia, IAG, Universidade de São Paulo, Rua do Matão 1226, São Paulo 05508-090, Brazil*

## ABSTRACT

The lithium present in the photospheres of solar-type stars is transported to the inner parts by convection, reaching regions even somewhat below the convection zone, by non-standard transport mechanisms. In stars with deeper convective zones, this element can reach regions with temperatures sufficient enough to be destroyed, implying in a lower Li content. More metallic stars show a deepening of their convective zones, so they could deplete more Li in comparison with stars of lower metallicity. In order to verify this effect and its amplitude, we selected stars with $\sim1\,\mathrm{M_\odot}$ and metallicities within a factor of two relative to the Sun. We studied a sample of 41 metal-rich and -poor solar analogues, and carried out a joint analysis with a sample of 77 solar twins from our previous work, resulting in a total sample of 118 stars covering the metallicity range $-0.3 \leq [\mathrm{Fe/H}] \leq +0.3$ dex. We employed high-resolution ($R = 115\,000$) and high-signal-to-noise ratio ($S/N = 400$–$1000$) HARPS spectra and determined the atmospheric parameters using a line-by-line differential analysis and the Li abundance through spectral synthesis. The ages and masses of the whole sample were improved by refining the isochronal method. We also investigated the impact of planets on Li. We found robust anticorrelations between Li abundance and both metallicity and age, with a significance above $10\sigma$ in both cases. Our results agree qualitatively with theoretical predictions and are useful to constrain non-standard models of Li depletion, and to better understand transport and mixing mechanisms inside stars.

## 1 INTRODUCTION

Although the stellar interior cannot be directly observed, the study of lithium in stars can provide important information about transport mechanisms near the base of the convective layer of the star, as this chemical element is destroyed by proton capture at relatively low temperatures, close to $2.5 \times 10^6$ K. However, the temperature at the base of the convective zone of solar-type stars at the main-sequence (MS) stage is $\sim2 \times 10^6$ K, according to standard models of stellar interiors, thus not reaching a temperature high enough to burn Li.

Such models (which do not consider effects of rotation, magnetic fields, chromospheric activity and star-spots, diffusion, and processes of mass-loss and mass accretion) predict that stars with masses similar to the Sun deplete some of their Li content during the pre-main sequence (PMS; Randich & Magrini 2021 and references therein). At this stage, the convective zone of the stars is deeper, enabling Li to go deep into the stellar interior, where it can be destroyed. The depletion should smoothly increase with lower masses and higher ages and metallicities (Cummings et al. 2017; Jeffries et al. 2021 and references therein). As the convective zones of the stars become shallower at the MS, it is not expected that they continue to destroy Li in their interiors.

For MS stars, Li depletion must be driven by other processes besides convection that are not included in the standard models. However, a consensus on the main mechanisms causing Li depletion on the PMS and MS phases has not yet been reached. There must exist a way to transport the original Li from the convective region to somewhat below the bottom of the convective zone.

In the standard model, as Li cannot reach the innermost parts of the Sun that are hot enough, its abundance stays constant over time (Dantona & Mazzitelli 1984). This is against observations of a deeper destruction of this element in older stars (Baumann et al. 2010; do Nascimento J. D. et al. 2013; Meléndez et al. 2014a; Carlos, Nissen & Meléndez 2016; Carlos et al. 2019). On the other hand, non-standard models, which include modelling of stellar rotation or extra transport mechanisms in the stellar interior, can predict some destruction of Li with time (Charbonnel & Talon 2005; do Nascimento et al. 2009; Xiong & Deng 2009; Denissenkov 2010; Andrássy & Spruit 2015), as it would be transported from the convective zone to the innermost regions.

The region where the extra transport mechanism would occur must be relatively shallow, immediately below the base of the convective zone, since the greater the depth in the stellar interior, the greater the temperature and, consequently, the destruction of Li, due to the exponential dependence of the nuclear reaction rates with temperature. If Li was transported to regions much deeper than the base of the convection zone, it would be quickly destroyed. This is contrary to observations of the slowly decreasing abundance of Li over time (Monroe et al. 2013), suggesting that the transport should not reach very deep. This is reinforced by the roughly constant Be abundances for solar twins of different stellar ages (Tucci Maia et al. 2015), as otherwise Be would also be depleted in older stars.

There are several mechanisms related to enhanced or suppressed Li depletion in solar-like stars (see e.g. Mishenina et al. 2020; Dumont et al. 2021a). At the PMS stage, enhanced magnetic activity and/or large star-spots coverage lead to an increase of the radii of cool dwarfs (Somers & Pinsonneault 2015a; Bouvier et al. 2016). Such radius inflation could produce a delayed Li depletion, at fixed mass and age, due to a significant reduction of the temperature at the base of the convective zone (Somers & Pinsonneault 2015b; Jeffries et al. 2017; Franciosini et al. 2022). This effect could also raise a dispersion of Li abundances at a fixed temperature during the PMS, especially in K dwarfs. Besides, distinct physical conditions in the stellar interior and episodic accretion could enhance Li depletion at this stage. Baraffe & Chabrier (2010) found that repeated accretion bursts during the early PMS could modify the internal structure of young low-mass stars in a manner that the central temperature is increased, thus increasing the Li depletion.

★ E-mail: giumartos@gmail.com

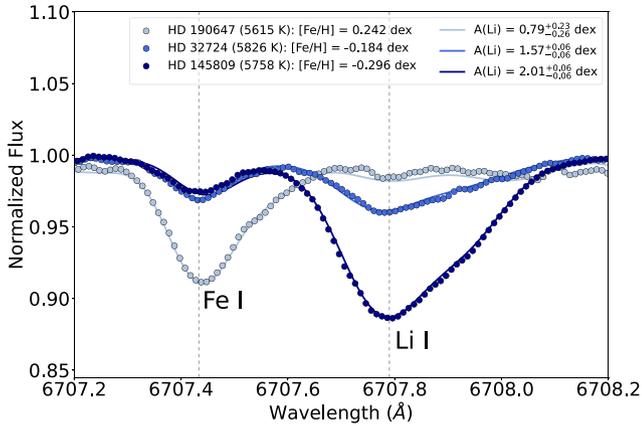

**Figure 1.** Spectra of stars of similar masses and same ages (within the uncertainties) HD190647, HD32 724, and HD145 809 around the Li line at 6707.8 Å. The line at 6707.4 Å is Fe I. The circles represent the observed spectrum and the solid line is the synthetic spectrum fit.

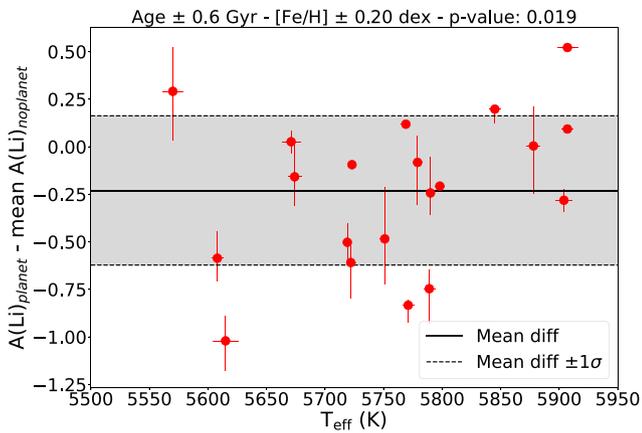

**Figure 2.** Difference in Li abundances between stars with planets and stars with no planets. The A(Li) of each planet-host was compared to the mean A(Li) of twin stars with no planets, with masses, ages, and [Fe/H] within $\pm 0.1\ M_{\odot}$, $\pm 0.6$ Gyr, and $\pm 0.20$ dex, respectively. The solid line represents the mean value of the difference, and the dashed lines represent the mean value $\pm 1\sigma$ (standard deviation).

Regarding the MS, the sources for Li depletion could be: mass-loss, where Li is diluted in Li-free material and removed from the atmosphere (Hobbs, Iben & Pilachowski 1989), and many processes that transport Li to the hotter regions of the stars, such as atomic diffusion, mainly on hotter stars that have shallow convective zones

(Michaud 1986); rotationally induced mixing (Chaboyer, Demarque & Pinsonneault 1995; Charbonnel & Talon 2005; Nascimento et al. 2009); and extra mixing at the base of the convection zone through overshooting (Schlattl & Weiss 1999; Xiong & Deng 2009; Baraffe et al. 2017) and rotation-induced mixing and internal gravity waves (Charbonnel & Talon 2005).

Another reason for Li depletion may be the formation and presence of planets around the stars (Israelian et al. 2009; Delgado Mena et al. 2014), but this claim has been disputed (Baumann et al. 2010). Nevertheless, it is intriguing that the Sun seems to have a Li abundance lower than solar twins of similar age (Carlos et al. 2019).

The Li abundance is also related to planet engulfment events that can lead to the detection of a greater amount of Li than expected (Meléndez et al. 2017). This event becomes clearer in the study of binary stars with distinct surface chemical composition (Gratton et al. 2001; Oh et al. 2018; Galarza et al. 2021). As these stars are formed from the same cloud and approximately at the same epoch, it is expected that both of them have similar abundance patterns. However, there exist systems in which one of the companions presents a greater abundance of refractory elements, including Li, suggesting that a planet has been engulfed in the past (Spina et al. 2021 and references therein).

Stellar rotation and activity can also influence the observed Li abundance. Takeda et al. (2010) found that the surface Li content in solar-type stars progressively decreases as the rotation rate decreases, i.e. slow-rotators stars deplete more Li than fast-rotators. As they found that the rotation of stars depends on activity, they also concluded that Li depletion depends on stellar activity. Beck et al. (2017) found a similar trend. Along the same lines, Denissenkov (2010) suggests that there is an increase in Li depletion in stars with stronger magnetic activity.

A prediction of the non-standard model by Castro et al. (2008) is an increase in the depletion of Li in metal-rich stars, due to the deepening of the convective zone in stars with higher metal content, as they present increased opacity. This theoretical prediction was verified only tentatively by the work of Carlos et al. (2016, 2019), who found a dependence of Li depletion with metallicity, but with a significance of only $2\sigma$. That may have occurred because the sample studied was composed of solar twins, with metallicities very similar to that of the Sun. Therefore, in order to obtain more evidence for this prediction, we studied a broader range of metallicities, including stars more and less metallic than the Sun. The results obtained here could be used to develop improved models of the stellar interior.

## 2 SAMPLE AND DATA

The sample studied was composed of 41 solar analogues, selected from an updated catalogue of stellar parameters from Ramírez & Meléndez (2005) based on the following conditions: $0.95 \le M \le 1.07$ ($M_{\odot}$), $5500 \le T_{\mathrm{eff}} \le 5900$ (K), and $4.1 \le \log g \le 4.6$ (dex), with metallicities $-0.3 \le$ [Fe/H] $\le -0.15$ and $+0.1 \le$ [Fe/H] $\le +0.34$ (dex). To cover the range of metallicities between $-0.15$ and $+0.1$, we included in the analysis the 77 solar twins studied by Spina et al. (2018) and Carlos et al. (2019).

We used spectra from the HARPS spectrograph, which are available at European Southern Observatory's online data base.[1] These spectra have high resolving power ($R = 115\,000$), and we selected those with individual signal-to-noise ratio (S/N) > 40. In order to obtain a very high S/N, several of these spectra were combined with IRAF. As a result, we obtained combined spectra with $S/N_{\text{total}} > 400$, approaching $S/N_{\text{total}} \sim 1000$ when possible. Before the combination, the spectra were corrected for the radial velocity of the star. Finally, the combined spectra were normalized using IRAF. Each spectrum was divided in seven segments, to improve the normalization process, and the continuum was normalized using polynomials of similar orders for each segment.

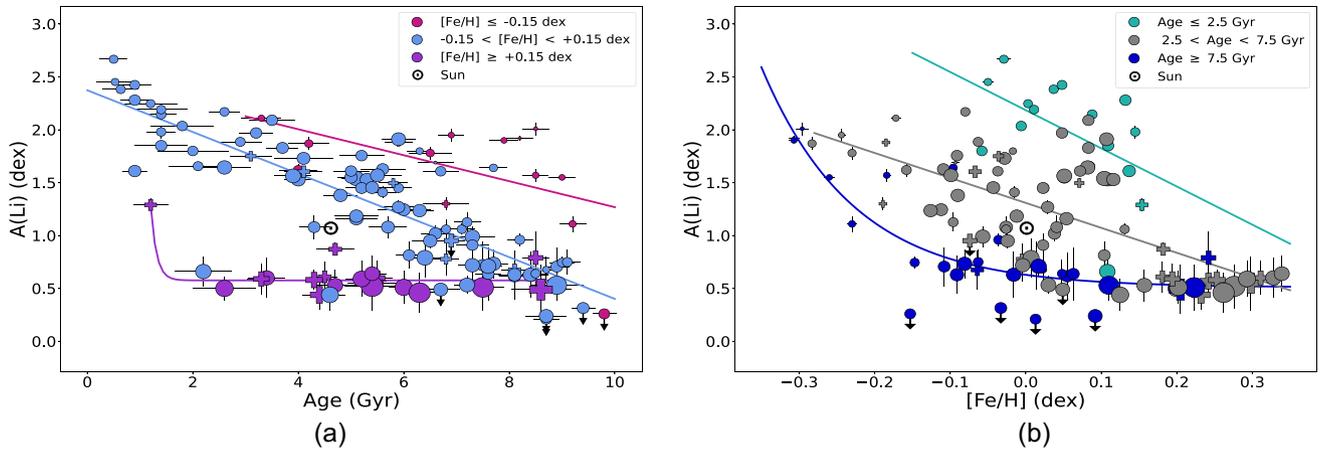

**Figure 3.** Illustration of the behaviour of NLTE Li abundance as a function of (a) age and (b) [Fe/H]. The size of the markers is proportional to the mass of the convective zone. Data with arrows pointing downwards correspond to upper limits of Li abundance. The crosses represent planet-host stars, and the circles stars without planets. The solid lines are fits to the points in the corresponding colour intervals of [Fe/H] and age, respectively.

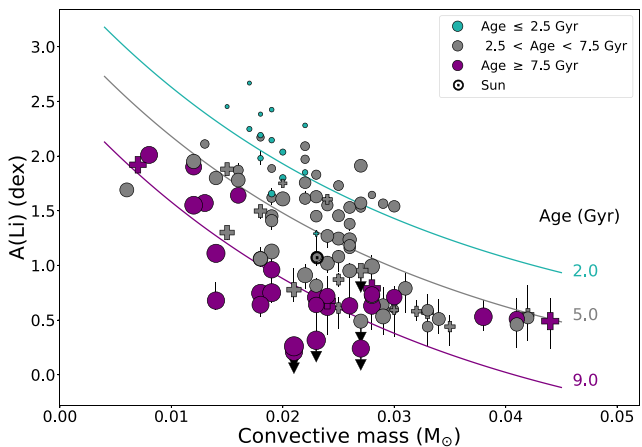

**Figure 4.** Li abundance as a function of convective mass. The lines are the fit to the exponential decrease in Li. The size of the markers is proportional to the age of the stars. Age intervals are separated by different colours. Data with arrows pointing downwards correspond to upper limits of Li abundance. The crosses represent planet-host stars, and the circles stars without planets.

Spectra taken before and after the HARPS upgrade that occurred in 2015 June were treated separately and then combined at the end of the correction processes. This is because the continuum of the spectra has a different shape due to the change of the optical fibres of the instrument. We also selected spectra from different dates, because as the Earth rotates around the Sun, the telluric lines will be located in different positions in the star's spectrum, being treated as outliers in the combining algorithm, thereby the merged spectrum suffers less telluric contamination.

# 3 STELLAR PARAMETERS

In order to acquire precise atmospheric parameters, such as effective temperature ($T_{\text{eff}}$), surface gravity (log $g$), metallicity ([Fe/H]), and microturbulence velocity ($v_t$), we carried out a line-by-line spectroscopic differential analysis, which consists in determining the values of these parameters in relation to a standard object with characteristics similar to the object of interest. As we are studying solar analogues, the standard object of comparison adopted was the Sun.

Seeking precise stellar parameters, it is necessary to have precise measurements of the equivalent widths of the iron lines, which was possible because we employed high quality spectra. Furthermore, the iron lines that were selected (Meléndez et al. 2014b) are relatively clean and include a broad range of excitation potentials, and at given excitation potential different line strengths are covered, to minimize the degeneracy of the derived stellar parameters. We measured the equivalent widths of 91 Fe I and 18 Fe II lines with IRAF, using a differential technique (Bedell et al. 2014) that consisted in comparing the continuum and line profile of the spectrum of the star and the Sun, so that the measurements are done strictly in the same way for each individual line, resulting in precise differential abundances relative to the Sun. We used mainly the asteroid Vesta, obtained also with the HARPS spectrograph, as a source of the reflected solar flux spectrum. We verified that this spectrum is cleaner from telluric lines. In a few cases the reference solar spectrum was obtained from a HARPS Moon spectrum, when the lines of interest were cleaner (than the Vesta spectrum) from telluric features. We adopted a Gaussian shape profile for the lines, and performed a deblending process when necessary.

The automatic code q2 (Ramírez et al. 2015) was used to obtain the atmospheric parameters of the 41 solar analogues based on the spectroscopic equilibrium, through the ionization and excitation balances. q2 uses the local thermodynamic equilibrium (LTE) code MOOG (Sneden et al. 2012) and was configured to employ the Kurucz model atmospheres (Castelli & Kurucz 2004).

The ages and masses were determined consistently in the same manner, both for the solar analogues and were revised for the solar twins, using a method similar to Spina et al. (2018), but with improved global metallicities. In their work, they employed Yonsei−Yale isochrones (Yi et al. 2001; Demarque, Kim & Yi 2004), with a shift of −0.04 dex in metallicity, so that for the Sun's stellar parameters the resulting mass and age coincide with the solar values, and another offset ($+\log_{10}(0.64 \times 10^{[\alpha/Fe]} + 0.36)$), to consider the contribution of alpha ($\alpha$) elements to the global metallicity of the star. That is because in stars with non-solar abundance ratios, in particular with $[\alpha/Fe] \neq 0$, the ratio [Fe/H] does not truly represent the total metallicity in the stellar interior. This way, the corrected metallicity would be given by equation (1) (Salaris, Chieffi & Straniero 1993), where Mg was adopted as the representative $\alpha$ element. The Mg abundances were obtained using the code $q2$, from equivalent widths of Mg I lines measured line-by-line with IRAF, using the line list of Meléndez et al. (2014b):

$$[M/H] = [Fe/H] - 0.04 + \log_{10}\left(0.64 \times 10^{[Mg/Fe]} + 0.36\right). \quad (1)$$

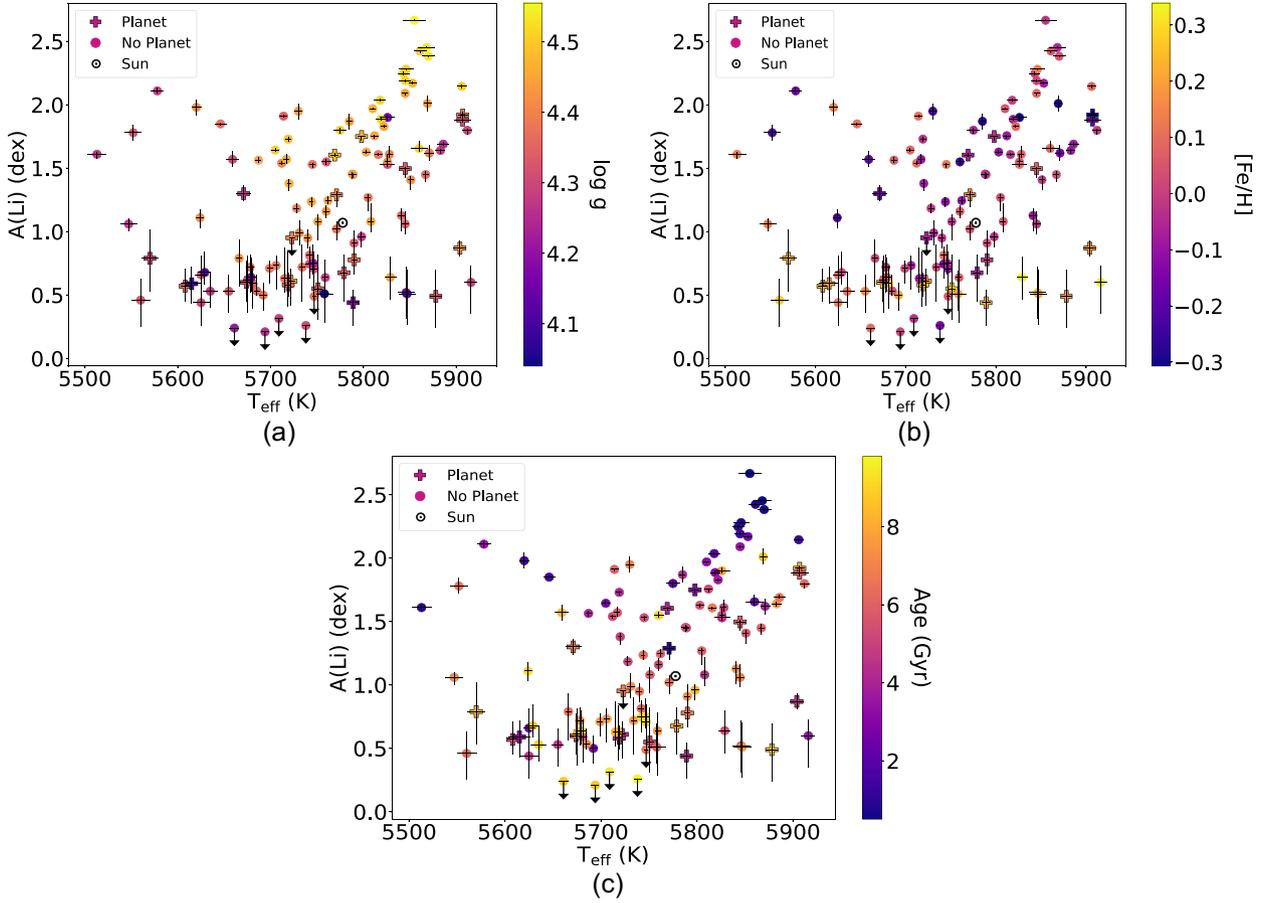

**Figure 5.** Behaviour of Li abundance with effective temperature ($T_{eff}$), colour coded by (a) surface gravity (log $g$); (b) metallicity ([Fe/H]); and (c) age. Stars with planets are represented by cross markers. Data with arrows pointing downwards correspond to upper limits of Li abundance.

However, for stars with [Fe/H] < 0, the Yonsei-Yale isochrones implemented in $q2$ already consider an enrichment in $\alpha$ elements, to follow the trend of galactic chemical evolution. Thus, using equation (1), we would consider doubly the $\alpha$ abundance for metal-poor stars. This way, it is necessary to remove the intrinsic $\alpha$-contribution from the isochrones used by $q2$, which have $[\alpha/Fe]$ as adopted in Meléndez et al. (2010): $[\alpha/Fe] = 0$ for $[Fe/H] \geq 0$, $[\alpha/Fe] = -0.3 \times [Fe/H]$ for $-1 < [Fe/H] < 0$, and $[\alpha/Fe] = +0.3$ for $[Fe/H] \leq -1$. Thus, the final global metallicity of the star is given by equations (2) and (3), for stars with $-1 < [Fe/H] < 0$ and $[Fe/H] \leq -1$, respectively. As for stars with $[Fe/H] \geq 0$ the ratio $[\alpha/Fe] = 0$ for the Yonsei–Yale isochrones used by $q2$, their final global metallicity is already given in equation (1).

$$[M/H] = [Fe/H] - 0.04 + \log_{10}\left(0.64 \times 10^{[Mg/Fe]} + 0.36\right)$$
$$- \log_{10}\left(0.64 \times 10^{-0.3 \times [Fe/H]} + 0.36\right). \quad (2)$$

$$[M/H] = [Fe/H] - 0.04 + \log_{10}\left(0.64 \times 10^{[Mg/Fe]} + 0.36\right)$$
$$- \log_{10}\left(0.64 \times 10^{+0.3} + 0.36\right). \quad (3)$$

The age and mass calculations were made using the code $q2$, adopting the method which uses only parallaxes, and not parallaxes together with log $g$, as made in Spina et al. (2018). This is because we used high-quality *Gaia* DR3 (Gaia Collaboration 2022) parallaxes, while Spina et al. (2018) employed *Gaia* DR1 (Lindegren et al. 2016) and even some old Hipparcos data. This way we avoid the use of the spectroscopic log $g$, which may have some degree of degeneracy with other atmospheric parameters. The adopted result was the most probable value (or the average value, when a most probable value was not available) of the ages/masses probability distribution given by $q2$, and the errors are the difference between the most probable value and the values considering $\pm 1\sigma$.

The parameters we obtained for the solar analogues are in excellent agreement with previous results from the literature, with mean differences and standard deviation for $T_{eff}$, log $g$, and [Fe/H] of $(11 \pm 14)$ K, $(0.02 \pm 0.01)$ dex, and $(0.015 \pm 0.011)$ dex, respectively. Regarding the ages and masses of solar twins, for which we redetermined their values, the mean difference and standard deviation between the values from Spina et al. (2018) and the new ones are $(0.43 \pm 0.43)$ Gyr and $(0.007 \pm 0.011)$ $M_\odot$.

The convective masses were obtained with the code Terra (Galarza, Meléndez & Cohen 2016). This code interpolates the values of the star's metallicity and mass with theoretical values (Spada et al. 2017) of the dependence of convective mass with mass and metallicity for solar-type stars.

## 4 LITHIUM ABUNDANCES

The Li abundances were calculated using the spectral synthesis method, employing the radiative transfer code MOOG, which uses a model atmosphere and a list of spectral lines around the characteristic line of the element of interest to solve the radiative transfer equation and generate a synthetic spectrum for the star. We synthesized the $^7$Li line at 6707.8 Å and nearby blends, characterizing manually the absorption feature for all stars. The macroturbulence velocity and projected velocity values required by MOOG were obtained using the relations found by dos Santos et al. (2016), and we adopted Kurucz model atmospheres.

The Li abundances as well as the abundances of the chemical species that interfere in the region around the Li line, such as $C_2$, CN, and Fe, were varied several times with the aim of minimizing the deviations between the synthetic and the observed spectra of the star. That is, until the Li line fit residuals were the minimum possible. Examples of these fits are shown in Fig. 1, where the circular markers represent the observed spectrum and the solid line is the spectral synthesis.

The errors associated with the measured abundances are observational ($\sigma_{obs}$) and systematic ($\sigma_{sys}$), being mainly dominated by the observational uncertainties, due to the quality of the spectrum and how deep the absorption feature is. The systematic uncertainties were determined by the errors of the atmospheric parameters. The total error was estimated as the quadratic sum of both error components: $\sigma_{A(Li)}^{-,+} = \sqrt{\sum \sigma_{sys}^2 + (\sigma_{obs}^{-,+})^2}$. The signs '−' and '+' correspond to the inferior and superior errors. That is, the difference between the best value found for the Li abundance and the minimum value, and the difference between the maximum and best values found, respectively.

The code MOOG assumes that the line formation in the photosphere occurs under LTE. Although it is a good first approximation, the lines are actually formed in a non-local thermodynamic equilibrium (NLTE) scenario. Therefore, we performed NLTE corrections employing the INSPECT data base[2] (Lind, Asplund & Barklem 2009), which presents both LTE and NLTE abundances for a grid of atmospheric parameters. In this way, it is possible to interpolate a NLTE correction (NLTE–LTE abundance) and add it to the LTE value obtained with MOOG, to infer the NLTE Li abundance.

## 5 RESULTS AND DISCUSSION

We carried out a joint analysis of the 77 solar twins from the work of Carlos et al. (2019) with the 41 solar analogues of our sample, resulting in a total sample of 118 stars. The solar values we adopted were: $A(Li)_\odot = 1.07^{+0.03}_{-0.02}$ dex, $T_{eff} = 5777$ K, and convective mass of $0.023$ $M_\odot$.

In their work, Carlos et al. (2019) adopted the values of the parameters from Spina et al. (2018). As explained in Section 3, we further improved the determination of masses and ages, but the differences are minor in relation to the values computed by Spina et al. (2018). Their Li abundances were also obtained using the same procedure as in our work, through the synthesis of the 6707.8 Å $^7$Li spectral feature, using the same line list and spectral synthesis technique. Therefore, our results are on the same scale as Carlos et al. (2019), but we recalculated the ages and masses of their sample, albeit only minor differences were found. The values obtained for the stellar parameters and Li abundances of the solar analogues are presented in Table A1. The parameters and updated ages and masses of the solar twins are given in Table A2.

Since the formation and presence of planets can influence the Li content of the stars, we searched The Extrasolar Planets Encyclopaedia[3] and found that 20 of the stars analysed are planet-hosts (14 from the sample of this study, and six from Carlos et al. 2019). These stars are represented with cross markers on the plots. We searched for differences in Li abundance due to the presence of planets by comparing the Li content of planet-hosts with the mean A(Li) of non-planet hosting 'twin' stars with similar masses, ages, and [Fe/H]. We found that stars with planets have depleted, in average, −0.23 dex of their Li content, with a standard deviation of 0.39 dex, in comparison with other stars within ±0.1 $M_\odot$ in mass, ±0.6 Gyr in age and ±0.20 dex in [Fe/H] (see Fig. 2). We performed a *t*-test, in which the null hypothesis was that there is no difference in A(Li) for stars with and without planets (expected mean value for the difference is zero), and found a significance of 98.1 per cent for our result. If we consider the interval of 0.10 or 0.15 dex for [Fe/H], the systematic decrease in Li is less significant (∼95 per cent).

Israelian et al. (2009) found systematically lower Li abundances in stars with planets around solar $T_{eff}$, relative to non-planet hosts, but that could be due to a bias, as only $T_{eff}$ was considered in the comparison, ignoring the effects due to age (log $g$) or metallicity ([Fe/H]). Indeed, when a more detailed comparison of the same sample is made between twin stars with and without planets, considering both similar $T_{eff}$, log $g$, and [Fe/H], no significant differences are found between both samples (Baumann et al. 2010). Different studies have been performed on the possible relation between Li and planets, and overall the results are inconclusive (e.g. Ghezzi et al. 2010; Gonzalez, Carlson & Tobin 2010; Ramírez et al. 2012; Bensby & Lind 2018; Pavlenko et al. 2018). Thus, it would be important to study a larger sample of planet-hosting stars and a comparison sample, to acquire more conclusive results.

We analysed the dependence of Li abundances with both [Fe/H] and age, and we found the correlation given by equation (4), which was obtained using the least-squares method, with the library LMFIT from PYTHON. Although the linear function is not a good representation for all the data points, the main objective of the fit was to evaluate the level of correlation between the parameters, and not to provide an exact model of stellar evolution.

$$A(Li) = -(2.70 \pm 0.22) \times [Fe/H] - (0.19 \pm 0.01) \times age$$
$$+ (2.36 \pm 0.08). \tag{4}$$

In the case of solar metallicity ([Fe/H] = 0), equation (4) simplifies to equation (5), which presents an anticorrelation with a significance of $14\sigma$ with age. This result is in line with the work of Carlos et al. (2019), who obtained the correlation A(Li) = (−0.20 ± 0.02) age + (2.44 ± 0.10), which also indicates de-depletion of Li with age, but with a significance of $10\sigma$.

$$A(Li) = -(0.19 \pm 0.01) \times age + (2.36 \pm 0.08). \tag{5}$$

---

They also studied a possible correlation of their residuals with metallicity, finding a significance of $2\sigma$ with [Fe/H], while we found a significance of $12\sigma$. Thus, our work demonstrates the existence of strong anticorrelations with both age and metallicity. This strong Li depletion is not predicted by the standard models of stellar interiors, as explained in Section 1. However, the depletion of Li with age and metallicity has already been suggested and found by other works, such as Ramírez et al. (2012) and Carlos et al. (2019).

Bensby & Lind (2018) found a potential Li-age correlation for the solar twins in their sample. This less clear result can be explained by the larger scatter in their Li abundances, as their spectra presented lower S/N ratio compared to our study, resulting in larger uncertainties in the stellar parameters. Randich & Magrini (2021) reported that Li depletion with age is indeed at work for MS stars, with a continuum depletion up to only 1 Gyr. This is against our results of Li depletion in the solar analogues with ages higher than this value. However, note that the stars of the clusters analysed by Randich & Magrini (2021) are not all solar analogues and solar twins, which are the focus of this study.

Fig. 3 shows the Li depletion behaviour with age and metallicity. From this figure it is clear the overall decrease in Li abundance with age and [Fe/H]. More metallic stars with larger convective masses (larger markers) present lower Li content. This depletion according to metallicity is qualitatively in line with the works of Castro et al. (2008) and Dumont et al. (2021b).

In order to illustrate the trend with convective mass, we show an exponential fit in Fig. 4, calculated using the function `optimize.curve_fit` from the PYTHON library SCIPY. It is possible to note the Li depletion with increasing convective masses and ages. At a given convective mass, younger stars (smaller markers) show higher A(Li), while older stars are systematically more depleted. The significance found for the exponential fit with convective mass is $2\sigma$.

The dependence of Li depletion with metallicity can be discerned even by visual inspection of spectra around the Li feature, as shown in Fig. 1. We plot the spectra of three stars of the sample around the 6707.8 Å Li feature near the iron (Fe I) line at 6707.4 Å. These stars have the same age, so we can discern the influence of metallicity on the observed Li abundances. It is possible to see that the higher the metal content (deeper Fe I line), the shallower the Li feature is (lower A(Li)). The impact of [Fe/H] on the depletion is reinforced by analysing the mass of the star. It is expected that low-mass stars deplete more Li in their interior, due to the deepening of the convective zone. Although HD 190 647 is the most massive star of the comparison in Fig. 1, it presents the lower Li content, due to its high metallicity.

Despite the small range of effective temperature and surface gravity considered in this study, we analysed a possible dependence of Li abundance with these parameters, since it is expected that stars with lower $T_{\text{eff}}$ deplete more Li than hotter stars, due to the deeper convective region. In Fig. 5, we plot A(Li) versus $T_{\text{eff}}$, colour coded by log $g$, [Fe/H] and age. From the plots, there is no clear dependence of A(Li) with $T_{\text{eff}}$, although it is possible to note a slight increase in the abundance towards higher temperatures, but not with a well-defined behaviour. The increase in the Li content with $T_{\text{eff}}$ agrees with the results of Bensby & Lind (2018), although in their work they considered a wider range of values.

The distribution of A(Li) with log $g$ is quite diverse, with a few Li-rich stars at high log $g$ and scatter at different log $g$ values. The higher Li abundance for higher log $g$ is expected, since those stars are typically younger, implying higher Li abundances. A similar complex correlation is shown for [Fe/H], but with more metal-rich stars at low Li abundances. It is also noted the depletion of Li with age, as the stars on the top of the corresponding plot are the younger ones. The few young stars with low Li at the bottom of the plot are, coincidentally, planet hosts.

Thus, the correlations of A(Li) are not well defined in $T_{\text{eff}}$ and log $g$ space. In Fig. 1, the cooler star is indeed the more Li-depleted. However, this behaviour is not valid for the hotter star, which is not the Li-richer one. As mentioned above, for this star the Li abundance seems to be more influenced by metallicity, as it is the more metallic among the ones shown in Fig. 1.

The stars with planets are more concentrated on the bottom of the plots, in agreement which the fact that planet-hosts of the sample are systematically depleted, as shown in Fig. 2. We found no relation between the presence of planets and the atmospheric parameters.

Finally, we checked for trends in the results due to the kinematics of the stars in the Galaxy, by calculating their galactic space velocities and inferring the probability to belong to a given stellar population (Reddy, Lambert & Allende Prieto 2006). We found that ~80 per cent of the stars have a probability >90 per cent of being from the thin disc, and only eight of them (the older ones) have a probability >50 per cent of being from the thick disc. Therefore, we can conclude that there is no significant trend in the results due to the location of stars, since most of them are in the thin disc of the Milky Way.

## 6 CONCLUSIONS

We used high-quality HARPS spectra of 41 solar analogues with metallicities up to a factor of two higher ([Fe/H] = +0.3) and lower ([Fe/H] = −0.3) than solar, to obtain their Li abundances via spectrum synthesis. These spectra had $S/N_{\text{total}}$ = 400–1000 and high resolving power ($R$ = 115 000). We obtained the atmospheric parameters using a line-by-line spectroscopic differential analysis and Li abundances were obtained through spectral synthesis. We calculated the ages and masses of the solar analogues of our sample and also of the solar twins from previous works (Spina et al. 2018; Carlos et al. 2019) using an isochronal method and taking into account the metallicity enhancement due to the contribution by the α elements.

After a joint analysis of our sample together with the solar twins, we found correlations of Li depletion with age and metallicity, with significances of $14\sigma$ and $12\sigma$, respectively, showing that older and more metallic stars present a smaller Li content. A possible explanation is that, in these stars, Li is burnt in the stellar interior with time during the MS (we did not considered stars in the PMS phase), and more metallic stars present deeper convective zones, making easier to take Li to the innermost and hotter parts, where it can be destroyed.

Qualitatively, our strong anticorrelation with age is in line with non-standard models of Li depletion (Charbonnel & Talon 2005; Nascimento et al. 2009; Xiong & Deng 2009; Denissenkov 2010; Andrássy & Spruit 2015), as those models are constructed to reproduce the low Li observed in the solar photosphere. The strong anticorrelation with metallicity observed in our sample is also within the predictions of non-standard models, due to the deepening of the convection zone (Castro et al. 2008; Dumont et al. 2021b). However, there is a lack of detailed predictions of Li depletion both with age and metallicity, preventing more detailed comparisons with our results.

We also investigated the influence of planets on the Li abundance, and found that planet-hosts are systematically depleted in −0.23 dex in comparison with twin stars without planets. However, it is necessary to analyse a larger sample in order to obtain more conclusive results.

We hope our work will contribute to constrain non-standard stellar evolution models and we encourage the study of theoretical Li depletion models as a function of age, mass (or convective mass), and metallicity, in order to perform comprehensive comparisons with the observations presented herein.

## ACKNOWLEDGEMENTS

The study was financed by Fundação de Amparo à Pesquisa do Estado de São Paulo (FAPESP), under grants 2022/05833-0, 2020/12679-1, and 2018/04055-8. We thank the colleagues of our group SAMPA (Stellar Atmospheres, Planets and Abundances) who gave support and training on the techniques employed in the study.

## DATA AVAILABILITY

The spectra analysed in this study are publicly available at ESO's data base website.

## APPENDIX A: LITHIUM ABUNDANCES AND ATMOSPHERIC PARAMETERS

**Table A1.** Li abundance, mass, age, and atmospheric parameters of the solar analogues selected for this study.

| HD | HIP | A(Li) LTE (dex) | A(Li) NLTE (dex) | [Fe/H] (dex) | Convective mass (M$_\odot$) | Mass (M$_\odot$) | Age (Gyr) | $T_{\rm eff}$ (K) | log $g$ (dex) | $v_{\rm t}$ (km s$^{-1}$) |
|---|---|---|---|---|---|---|---|---|---|---|
| 100777 | 56572 | $0.45^{+0.27}_{-0.24}$ | $0.55^{+0.27}_{-0.24}$ | $0.298 \pm 0.010$ | 0.042 | $1.03^{+0.01}_{-0.01}$ | $5.40^{+0.44}_{-0.48}$ | $5552 \pm 9$ | $4.32 \pm 0.02$ | $0.88 \pm 0.03$ |
| 106116 | 59532 | $0.46^{+0.13}_{-0.17}$ | $0.53^{+0.13}_{-0.17}$ | $0.157 \pm 0.004$ | 0.029 | $1.04^{+0.01}_{-0.01}$ | $4.70^{+0.46}_{-0.37}$ | $5681 \pm 4$ | $4.34 \pm 0.01$ | $0.98 \pm 0.01$ |
| 108309 | 60729 | $1.01^{+0.04}_{-0.06}$ | $1.06^{+0.04}_{-0.06}$ | $0.131 \pm 0.005$ | 0.018 | $1.09^{+0.01}_{-0.01}$ | $7.10^{+0.21}_{-0.27}$ | $5789 \pm 5$ | $4.18 \pm 0.01$ | $1.13 \pm 0.01$ |
| 114729 | 64459 | $1.91^{+0.02}_{-0.02}$ | $1.92^{+0.02}_{-0.02}$ | $-0.307 \pm 0.007$ | 0.007 | $1.04^{+0.01}_{-0.02}$ | $8.20^{+0.26}_{-0.24}$ | $5847 \pm 9$ | $4.11 \pm 0.02$ | $1.34 \pm 0.02$ |
| 117105 | 65737 | $1.90^{+0.02}_{-0.02}$ | $1.90^{+0.02}_{-0.02}$ | $-0.307 \pm 0.006$ | 0.012 | $0.98^{+0.01}_{-0.01}$ | $7.90^{+0.31}_{-0.24}$ | $5907 \pm 9$ | $4.31 \pm 0.02$ | $1.25 \pm 0.02$ |
| 117207 | 65808 | $0.49^{+0.14}_{-0.12}$ | $0.57^{+0.14}_{-0.12}$ | $0.242 \pm 0.007$ | 0.033 | $1.05^{+0.01}_{-0.01}$ | $5.40^{+0.29}_{-0.26}$ | $5659 \pm 7$ | $4.30 \pm 0.02$ | $0.97 \pm 0.02$ |
| 134606 | 74653 | $0.51^{+0.17}_{-0.15}$ | $0.60^{+0.17}_{-0.15}$ | $0.311 \pm 0.008$ | 0.029 | $1.06^{+0.01}_{-0.01}$ | $5.50^{+0.29}_{-0.26}$ | $5625 \pm 9$ | $4.29 \pm 0.02$ | $0.97 \pm 0.02$ |
| 13612B | 10303 | $1.47^{+0.02}_{-0.02}$ | $1.53^{+0.02}_{-0.02}$ | $0.109 \pm 0.003$ | 0.027 | $1.03^{+0.01}_{-0.01}$ | $4.00^{+0.28}_{-0.28}$ | $5719 \pm 3$ | $4.36 \pm 0.01$ | $0.95 \pm 0.01$ |
| 140901 | 77358 | $0.59^{+0.15}_{-0.13}$ | $0.66^{+0.15}_{-0.13}$ | $0.108 \pm 0.004$ | 0.030 | $1.01^{+0.01}_{-0.01}$ | $2.20^{+0.80}_{-0.28}$ | $5624 \pm 4$ | $4.42 \pm 0.02$ | $1.00 \pm 0.01$ |
| 145809 | 79524 | $2.00^{+0.06}_{-0.06}$ | $2.01^{+0.06}_{-0.06}$ | $-0.296 \pm 0.008$ | 0.008 | $1.03^{+0.01}_{-0.01}$ | $8.50^{+0.28}_{-0.23}$ | $5758 \pm 9$ | $4.04 \pm 0.02$ | $1.25 \pm 0.01$ |
| 1461 | 1499 | $0.81^{+0.06}_{-0.06}$ | $0.87^{+0.06}_{-0.06}$ | $0.182 \pm 0.003$ | 0.025 | $1.07^{+0.01}_{-0.01}$ | $4.70^{+0.36}_{-0.35}$ | $5751 \pm 4$ | $4.33 \pm 0.01$ | $1.01 \pm 0.01$ |
| 161612 | 87116 | $0.42^{+0.10}_{-0.12}$ | $0.50^{+0.10}_{-0.12}$ | $0.200 \pm 0.007$ | 0.034 | $1.03^{+0.02}_{-0.01}$ | $2.60^{+0.29}_{-0.29}$ | $5620 \pm 5$ | $4.42 \pm 0.02$ | $0.85 \pm 0.02$ |
| 185615 | 96854 | $0.46^{+0.15}_{-0.13}$ | $0.53^{+0.15}_{-0.13}$ | $0.110 \pm 0.007$ | 0.038 | $0.97^{+0.02}_{-0.01}$ | $8.90^{+0.25}_{-0.75}$ | $5578 \pm 7$ | $4.28 \pm 0.02$ | $0.93 \pm 0.02$ |
| 190647 | 99115 | $0.71^{+0.23}_{-0.26}$ | $0.79^{+0.23}_{-0.26}$ | $0.242 \pm 0.009$ | 0.028 | $1.08^{+0.01}_{-0.01}$ | $8.50^{+0.30}_{-0.28}$ | $5615 \pm 11$ | $4.09 \pm 0.02$ | $1.07 \pm 0.02$ |
| 198075 | 102664 | $1.86^{+0.06}_{-0.06}$ | $1.87^{+0.06}_{-0.06}$ | $-0.283 \pm 0.005$ | 0.016 | $0.96^{+0.01}_{-0.01}$ | $4.20^{+0.35}_{-0.42}$ | $5829 \pm 6$ | $4.45 \pm 0.02$ | $1.10 \pm 0.01$ |
| 203384 | 105483 | $0.43^{+0.20}_{-0.21}$ | $0.52^{+0.20}_{-0.21}$ | $0.276 \pm 0.010$ | 0.042 | $1.02^{+0.01}_{-0.01}$ | $5.40^{+0.34}_{-0.71}$ | $5560 \pm 10$ | $4.31 \pm 0.02$ | $0.94 \pm 0.03$ |
| 203432 | 105606 | $0.51^{+0.23}_{-0.25}$ | $0.60^{+0.23}_{-0.25}$ | $0.328 \pm 0.007$ | 0.030 | $1.06^{+0.02}_{-0.01}$ | $3.40^{+0.30}_{-0.30}$ | $5646 \pm 6$ | $4.37 \pm 0.02$ | $1.01 \pm 0.02$ |
| 204313 | 106006 | $0.55^{+0.10}_{-0.19}$ | $0.61^{+0.10}_{-0.19}$ | $0.181 \pm 0.005$ | 0.025 | $1.07^{+0.02}_{-0.01}$ | $4.50^{+0.23}_{-0.26}$ | $5760 \pm 5$ | $4.32 \pm 0.01$ | $1.03 \pm 0.01$ |
| 207832 | 107985 | $1.23^{+0.03}_{-0.09}$ | $1.29^{+0.03}_{-0.09}$ | $0.154 \pm 0.004$ | 0.023 | $1.07^{+0.01}_{-0.01}$ | $1.20^{+0.25}_{-0.41}$ | $5730 \pm 4$ | $4.44 \pm 0.01$ | $1.01 \pm 0.01$ |
| 210277 | 109378 | $0.40^{+0.21}_{-0.25}$ | $0.49^{+0.21}_{-0.25}$ | $0.205 \pm 0.011$ | 0.044 | $0.98^{+0.02}_{-0.01}$ | $8.60^{+0.88}_{-0.25}$ | $5513 \pm 10$ | $4.26 \pm 0.03$ | $0.91 \pm 0.03$ |
| 211415 | 110109 | $1.77^{+0.06}_{-0.06}$ | $1.78^{+0.06}_{-0.06}$ | $-0.230 \pm 0.004$ | 0.016 | $0.98^{+0.01}_{-0.01}$ | $6.50^{+0.50}_{-0.31}$ | $5871 \pm 5$ | $4.38 \pm 0.02$ | $1.15 \pm 0.01$ |
| 212708 | 110843 | $0.51^{+0.14}_{-0.20}$ | $0.59^{+0.14}_{-0.20}$ | $0.293 \pm 0.008$ | 0.033 | $1.07^{+0.01}_{-0.01}$ | $5.20^{+0.30}_{-0.29}$ | $5671 \pm 8$ | $4.27 \pm 0.02$ | $1.00 \pm 0.02$ |
| 28185 | 20723 | $0.36^{+0.10}_{-0.18}$ | $0.44^{+0.10}_{-0.18}$ | $0.232 \pm 0.007$ | 0.035 | $1.04^{+0.02}_{-0.01}$ | $4.40^{+0.42}_{-0.38}$ | $5655 \pm 6$ | $4.32 \pm 0.02$ | $0.96 \pm 0.02$ |
| 30306 | 21731 | $0.42^{+0.19}_{-0.24}$ | $0.51^{+0.19}_{-0.24}$ | $0.223 \pm 0.008$ | 0.041 | $1.00^{+0.01}_{-0.01}$ | $7.50^{+0.39}_{-0.42}$ | $5547 \pm 9$ | $4.27 \pm 0.02$ | $0.89 \pm 0.02$ |
| 31527 | 22905 | $1.87^{+0.02}_{-0.02}$ | $1.88^{+0.02}_{-0.02}$ | $-0.185 \pm 0.004$ | 0.015 | $1.00^{+0.01}_{-0.01}$ | $5.90^{+0.28}_{-0.28}$ | $5907 \pm 5$ | $4.38 \pm 0.01$ | $1.15 \pm 0.01$ |
| 32724 | 23627 | $1.55^{+0.06}_{-0.06}$ | $1.57^{+0.06}_{-0.06}$ | $-0.184 \pm 0.005$ | 0.013 | $1.02^{+0.01}_{-0.01}$ | $8.50^{+0.24}_{-0.25}$ | $5826 \pm 8$ | $4.16 \pm 0.02$ | $1.22 \pm 0.01$ |
| 43834 | 29271 | $0.37^{+0.12}_{-0.18}$ | $0.44^{+0.12}_{-0.18}$ | $0.125 \pm 0.005$ | 0.033 | $1.00^{+0.01}_{-0.01}$ | $4.60^{+0.28}_{-0.28}$ | $5608 \pm 5$ | $4.37 \pm 0.01$ | $0.93 \pm 0.01$ |
| 47186 | 31540 | $0.50^{+0.10}_{-0.09}$ | $0.58^{+0.10}_{-0.09}$ | $0.253 \pm 0.007$ | 0.032 | $1.06^{+0.01}_{-0.01}$ | $4.30^{+0.37}_{-0.37}$ | $5674 \pm 6$ | $4.33 \pm 0.02$ | $0.97 \pm 0.02$ |
| 53705 | 34065 | $1.09^{+0.07}_{-0.08}$ | $1.11^{+0.07}_{-0.08}$ | $-0.230 \pm 0.006$ | 0.014 | $0.99^{+0.02}_{-0.01}$ | $9.20^{+0.25}_{-0.26}$ | $5826 \pm 8$ | $4.29 \pm 0.02$ | $1.17 \pm 0.02$ |
| 66221 | 39298 | $0.43^{+0.18}_{-0.13}$ | $0.51^{+0.18}_{-0.13}$ | $0.202 \pm 0.007$ | 0.034 | $1.03^{+0.01}_{-0.01}$ | $6.00^{+0.29}_{-0.29}$ | $5625 \pm 6$ | $4.32 \pm 0.02$ | $0.93 \pm 0.02$ |
| 66653 | 39330 | $1.94^{+0.06}_{-0.06}$ | $1.98^{+0.06}_{-0.06}$ | $0.145 \pm 0.004$ | 0.018 | $1.10^{+0.02}_{-0.01}$ | $1.40^{+0.22}_{-0.32}$ | $5869 \pm 4$ | $4.43 \pm 0.01$ | $1.07 \pm 0.01$ |
| 67458 | 39710 | $2.10^{+0.02}_{-0.02}$ | $2.11^{+0.02}_{-0.02}$ | $-0.172 \pm 0.006$ | 0.013 | $1.02^{+0.01}_{-0.01}$ | $3.30^{+0.39}_{-0.37}$ | $5904 \pm 7$ | $4.43 \pm 0.02$ | $1.13 \pm 0.02$ |
| 70642 | 40952 | $0.52^{+0.13}_{-0.16}$ | $0.59^{+0.13}_{-0.16}$ | $0.194 \pm 0.004$ | 0.030 | $1.05^{+0.01}_{-0.01}$ | $3.30^{+0.27}_{-0.26}$ | $5692 \pm 4$ | $4.37 \pm 0.01$ | $0.94 \pm 0.01$ |
| 72769 | 42011 | $0.55^{+0.16}_{-0.18}$ | $0.64^{+0.16}_{-0.18}$ | $0.339 \pm 0.008$ | 0.029 | $1.07^{+0.01}_{-0.01}$ | $5.40^{+0.28}_{-0.26}$ | $5635 \pm 8$ | $4.27 \pm 0.02$ | $0.98 \pm 0.02$ |
| 74014 | 42634 | $0.37^{+0.17}_{-0.21}$ | $0.46^{+0.17}_{-0.21}$ | $0.262 \pm 0.008$ | 0.041 | $1.02^{+0.02}_{-0.01}$ | $6.30^{+0.40}_{-0.44}$ | $5570 \pm 9$ | $4.29 \pm 0.02$ | $0.91 \pm 0.02$ |
| 76151 | 43726 | $1.80^{+0.01}_{-0.01}$ | $1.85^{+0.01}_{-0.01}$ | $0.109 \pm 0.004$ | 0.022 | $1.06^{+0.02}_{-0.01}$ | $1.40^{+0.44}_{-0.55}$ | $5785 \pm 3$ | $4.44 \pm 0.01$ | $1.01 \pm 0.01$ |
| 78612 | 44896 | $1.53^{+0.02}_{-0.02}$ | $1.55^{+0.02}_{-0.02}$ | $-0.260 \pm 0.007$ | 0.012 | $1.00^{+0.01}_{-0.01}$ | $9.00^{+0.27}_{-0.22}$ | $5846 \pm 9$ | $4.18 \pm 0.02$ | $1.25 \pm 0.02$ |
| 83529 | 47225 | $1.94^{+0.06}_{-0.06}$ | $1.95^{+0.06}_{-0.06}$ | $-0.244 \pm 0.005$ | 0.012 | $1.00^{+0.01}_{-0.01}$ | $6.90^{+0.27}_{-0.22}$ | $5916 \pm 7$ | $4.28 \pm 0.02$ | $1.23 \pm 0.01$ |
| 89454 | 50534 | $1.55^{+0.02}_{-0.03}$ | $1.61^{+0.02}_{-0.03}$ | $0.137 \pm 0.005$ | 0.024 | $1.06^{+0.01}_{-0.01}$ | $0.90^{+0.24}_{-0.24}$ | $5722 \pm 5$ | $4.45 \pm 0.01$ | $1.02 \pm 0.01$ |
| 95521 | 53837 | $1.59^{+0.06}_{-0.06}$ | $1.62^{+0.06}_{-0.06}$ | $-0.158 \pm 0.004$ | 0.019 | $0.98^{+0.01}_{-0.01}$ | $4.00^{+0.34}_{-0.36}$ | $5771 \pm 5$ | $4.42 \pm 0.01$ | $0.99 \pm 0.01$ |
| 96700 | 54400 | $1.28^{+0.06}_{-0.06}$ | $1.30^{+0.06}_{-0.06}$ | $-0.189 \pm 0.004$ | 0.015 | $1.00^{+0.01}_{-0.01}$ | $6.80^{+0.64}_{-0.24}$ | $5878 \pm 6$ | $4.33 \pm 0.02$ | $1.16 \pm 0.01$ |

**Table A2.** Li abundance, mass, age, and atmospheric parameters of the solar twins.

| HD | HIP | A(Li) LTE (dex) | A(Li) NLTE (dex) | [Fe/H] (dex) | Convective mass (M$_\odot$) | Mass (M$_\odot$) | Age (Gyr) | $T_{\rm eff}$ (K) | log $g$ (dex) | $v_t$ (km s$^{-1}$) |
|---|---|---|---|---|---|---|---|---|---|---|
| 13357A | 10175 | $1.69^{+0.01}_{-0.02}$ | $1.73^{+0.01}_{-0.02}$ | $-0.028 \pm 0.002$ | 0.025 | $0.98^{+0.01}_{-0.01}$ | $4.10^{+0.63}_{-0.48}$ | $5719 \pm 3$ | $4.49 \pm 0.01$ | $0.97 \pm 0.01$ |
| 13612B | 10303 | $1.49^{+0.01}_{-0.01}$ | $1.54^{+0.01}_{-0.01}$ | $0.104 \pm 0.003$ | 0.030 | $1.01^{+0.01}_{-0.01}$ | $5.10^{+0.28}_{-0.26}$ | $5712 \pm 3$ | $4.39 \pm 0.01$ | $1.08 \pm 0.01$ |
| 44665A | 30158 | $0.67^{+0.10}_{-0.20}$ | $0.72^{+0.10}_{-0.20}$ | $-0.004 \pm 0.003$ | 0.028 | $0.97^{+0.01}_{-0.01}$ | $7.30^{+0.29}_{-0.25}$ | $5678 \pm 4$ | $4.37 \pm 0.01$ | $1.05 \pm 0.01$ |
| 110537 | 62039 | $0.76^{+0.07}_{-0.18}$ | $0.81^{+0.07}_{-0.18}$ | $0.104 \pm 0.003$ | 0.023 | $1.05^{+0.01}_{-0.01}$ | $6.10^{+0.27}_{-0.22}$ | $5742 \pm 3$ | $4.34 \pm 0.01$ | $0.99 \pm 0.01$ |
| 11195 | 8507 | $1.53^{+0.04}_{-0.03}$ | $1.57^{+0.04}_{-0.03}$ | $-0.099 \pm 0.003$ | 0.027 | $0.95^{+0.01}_{-0.01}$ | $5.50^{+0.39}_{-0.55}$ | $5717 \pm 3$ | $4.46 \pm 0.01$ | $0.94 \pm 0.01$ |
| 114174 | 64150 | $0.44^{+0.00}_{-0.00}$ | $0.49^{+0.00}_{-0.00}$ | $0.049 \pm 0.003$ | 0.027 | $1.00^{+0.01}_{-0.01}$ | $6.70^{+0.34}_{-0.38}$ | $5747 \pm 2$ | $4.37 \pm 0.01$ | $1.04 \pm 0.01$ |
| 115031 | 64673 | $1.78^{+0.01}_{-0.04}$ | $1.80^{+0.01}_{-0.04}$ | $-0.017 \pm 0.004$ | 0.014 | $1.06^{+0.01}_{-0.01}$ | $6.30^{+0.23}_{-0.25}$ | $5912 \pm 5$ | $4.29 \pm 0.01$ | $1.05 \pm 0.01$ |
| 115169 | 64713 | $1.42^{+0.01}_{-0.04}$ | $1.45^{+0.01}_{-0.04}$ | $-0.043 \pm 0.003$ | 0.025 | $0.98^{+0.02}_{-0.01}$ | $5.40^{+0.40}_{-0.41}$ | $5788 \pm 4$ | $4.43 \pm 0.01$ | $1.01 \pm 0.01$ |
| 117126 | 65708 | $0.71^{+0.14}_{-0.09}$ | $0.75^{+0.14}_{-0.09}$ | $-0.063 \pm 0.005$ | 0.019 | $1.01^{+0.02}_{-0.01}$ | $9.10^{+0.24}_{-0.24}$ | $5746 \pm 5$ | $4.22 \pm 0.01$ | $1.13 \pm 0.01$ |
| 122194 | 68468 | $1.46^{+0.01}_{-0.07}$ | $1.50^{+0.01}_{-0.07}$ | $0.071 \pm 0.004$ | 0.018 | $1.07^{+0.01}_{-0.01}$ | $5.80^{+0.29}_{-0.25}$ | $5845 \pm 5$ | $4.33 \pm 0.01$ | $1.06 \pm 0.01$ |
| 12264 | 9349 | $2.01^{+0.01}_{-0.01}$ | $2.04^{+0.01}_{-0.01}$ | $-0.006 \pm 0.003$ | 0.020 | $1.02^{+0.02}_{-0.01}$ | $1.80^{+0.24}_{-0.94}$ | $5818 \pm 6$ | $4.51 \pm 0.01$ | $1.03 \pm 0.01$ |
| 124523 | 69645 | $1.04^{+0.06}_{-0.14}$ | $1.08^{+0.06}_{-0.14}$ | $-0.026 \pm 0.004$ | 0.025 | $0.98^{+0.01}_{-0.01}$ | $5.70^{+0.31}_{-0.41}$ | $5751 \pm 3$ | $4.43 \pm 0.01$ | $1.04 \pm 0.01$ |
| 129814 | 72043 | $1.03^{+0.10}_{-0.18}$ | $1.06^{+0.10}_{-0.18}$ | $-0.026 \pm 0.003$ | 0.018 | $1.03^{+0.02}_{-0.02}$ | $6.80^{+0.26}_{-0.29}$ | $5845 \pm 4$ | $4.34 \pm 0.01$ | $1.08 \pm 0.01$ |
| 131923 | 73241 | $0.18^{+0.00}_{-0.00}$ | $0.24^{+0.00}_{-0.00}$ | $0.092 \pm 0.005$ | 0.027 | $1.02^{+0.01}_{-0.01}$ | $8.70^{+0.28}_{-0.50}$ | $5661 \pm 5$ | $4.21 \pm 0.01$ | $1.12 \pm 0.01$ |
| 133600 | 73815 | $0.87^{+0.10}_{-0.12}$ | $0.91^{+0.10}_{-0.12}$ | $0.023 \pm 0.003$ | 0.022 | $1.02^{+0.01}_{-0.01}$ | $7.30^{+0.33}_{-0.29}$ | $5790 \pm 3$ | $4.33 \pm 0.01$ | $1.10 \pm 0.01$ |
| 134664 | 74389 | $2.06^{+0.01}_{-0.01}$ | $2.09^{+0.01}_{-0.01}$ | $0.083 \pm 0.003$ | 0.022 | $1.05^{+0.01}_{-0.01}$ | $3.50^{+0.45}_{-0.43}$ | $5845 \pm 3$ | $4.44 \pm 0.01$ | $0.99 \pm 0.01$ |
| 135101 | 74432 | $0.59^{+0.16}_{-0.11}$ | $0.64^{+0.16}_{-0.11}$ | $0.048 \pm 0.005$ | 0.018 | $1.06^{+0.01}_{-0.01}$ | $8.50^{+0.24}_{-0.25}$ | $5679 \pm 5$ | $4.17 \pm 0.01$ | $1.07 \pm 0.01$ |
| 138573 | 76114 | $0.91^{+0.06}_{-0.09}$ | $0.95^{+0.06}_{-0.09}$ | $-0.024 \pm 0.003$ | 0.026 | $0.98^{+0.01}_{-0.01}$ | $6.50^{+0.29}_{-0.42}$ | $5740 \pm 3$ | $4.41 \pm 0.01$ | $1.03 \pm 0.01$ |
| 140538 | 77052 | $1.51^{+0.02}_{-0.02}$ | $1.56^{+0.02}_{-0.02}$ | $0.051 \pm 0.003$ | 0.029 | $0.99^{+0.01}_{-0.01}$ | $3.90^{+0.35}_{-0.34}$ | $5687 \pm 3$ | $4.45 \pm 0.01$ | $1.07 \pm 0.01$ |
| 142331 | 77883 | $0.66^{+0.06}_{-0.11}$ | $0.71^{+0.06}_{-0.11}$ | $0.017 \pm 0.003$ | 0.030 | $0.97^{+0.01}_{-0.01}$ | $7.60^{+0.23}_{-0.29}$ | $5699 \pm 3$ | $4.38 \pm 0.01$ | $1.01 \pm 0.01$ |
| 145825 | 79578 | $1.94^{+0.01}_{-0.01}$ | $1.97^{+0.01}_{-0.01}$ | $0.048 \pm 0.003$ | 0.022 | $1.03^{+0.01}_{-0.01}$ | $3.20^{+0.31}_{-0.33}$ | $5810 \pm 3$ | $4.46 \pm 0.01$ | $0.96 \pm 0.01$ |
| 145927 | 79715 | $1.57^{+0.01}_{-0.09}$ | $1.61^{+0.01}_{-0.09}$ | $-0.037 \pm 0.004$ | 0.020 | $1.01^{+0.01}_{-0.01}$ | $6.70^{+0.35}_{-0.32}$ | $5816 \pm 4$ | $4.38 \pm 0.01$ | $1.10 \pm 0.01$ |
| 146233 | 79672 | $1.05^{+0.14}_{-0.08}$ | $1.08^{+0.14}_{-0.08}$ | $0.041 \pm 0.003$ | 0.023 | $1.02^{+0.01}_{-0.02}$ | $4.30^{+0.28}_{-0.29}$ | $5808 \pm 3$ | $4.44 \pm 0.01$ | $0.98 \pm 0.01$ |
| 150248 | 81746 | $0.59^{+0.24}_{-0.11}$ | $0.63^{+0.24}_{-0.11}$ | $-0.091 \pm 0.003$ | 0.026 | $0.96^{+0.01}_{-0.01}$ | $8.10^{+0.24}_{-0.27}$ | $5715 \pm 3$ | $4.37 \pm 0.01$ | $1.07 \pm 0.01$ |
| 153631 | 83276 | $1.67^{+0.02}_{-0.01}$ | $1.69^{+0.02}_{-0.01}$ | $-0.093 \pm 0.005$ | 0.006 | $1.12^{+0.01}_{-0.01}$ | $6.60^{+0.33}_{-0.34}$ | $5886 \pm 6$ | $4.24 \pm 0.01$ | $1.01 \pm 0.01$ |
| 157347 | 85042 | $0.48^{+0.12}_{-0.04}$ | $0.53^{+0.12}_{-0.04}$ | $0.030 \pm 0.003$ | 0.029 | $0.98^{+0.02}_{-0.01}$ | $7.20^{+0.23}_{-0.25}$ | $5685 \pm 3$ | $4.35 \pm 0.01$ | $0.91 \pm 0.01$ |
| 16008 | 11915 | $1.57^{+0.01}_{-0.01}$ | $1.60^{+0.01}_{-0.01}$ | $-0.067 \pm 0.004$ | 0.024 | $0.98^{+0.01}_{-0.01}$ | $4.10^{+0.36}_{-0.35}$ | $5769 \pm 4$ | $4.48 \pm 0.01$ | $0.95 \pm 0.01$ |
| 163441 | 87769 | $1.57^{+0.06}_{-0.04}$ | $1.61^{+0.06}_{-0.04}$ | $0.072 \pm 0.004$ | 0.022 | $1.04^{+0.01}_{-0.01}$ | $5.00^{+0.30}_{-0.38}$ | $5828 \pm 3$ | $4.40 \pm 0.01$ | $1.05 \pm 0.01$ |
| 167060 | 89650 | $1.38^{+0.02}_{-0.03}$ | $1.41^{+0.02}_{-0.03}$ | $-0.015 \pm 0.003$ | 0.019 | $1.02^{+0.01}_{-0.01}$ | $5.60^{+0.27}_{-0.22}$ | $5851 \pm 3$ | $4.42 \pm 0.01$ | $0.98 \pm 0.01$ |
| 183579 | 96160 | $1.72^{+0.01}_{-0.03}$ | $1.75^{+0.01}_{-0.03}$ | $-0.036 \pm 0.003$ | 0.020 | $1.01^{+0.01}_{-0.01}$ | $3.10^{+0.41}_{-0.37}$ | $5798 \pm 4$ | $4.48 \pm 0.01$ | $1.16 \pm 0.02$ |
| 183658 | 95962 | $1.23^{+0.02}_{-0.11}$ | $1.27^{+0.02}_{-0.11}$ | $0.029 \pm 0.003$ | 0.024 | $1.01^{+0.01}_{-0.01}$ | $5.90^{+0.28}_{-0.28}$ | $5805 \pm 3$ | $4.38 \pm 0.01$ | $1.10 \pm 0.01$ |
| 19467 | 14501 | $0.22^{+0.00}_{-0.00}$ | $0.26^{+0.00}_{-0.00}$ | $-0.153 \pm 0.003$ | 0.021 | $0.97^{+0.01}_{-0.01}$ | $9.80^{+0.19}_{-0.19}$ | $5738 \pm 4$ | $4.30 \pm 0.01$ | $1.04 \pm 0.01$ |
| 19518 | 14614 | $1.60^{+0.01}_{-0.01}$ | $1.63^{+0.01}_{-0.01}$ | $-0.109 \pm 0.004$ | 0.023 | $0.97^{+0.01}_{-0.01}$ | $5.60^{+0.38}_{-0.47}$ | $5803 \pm 4$ | $4.45 \pm 0.01$ | $0.99 \pm 0.01$ |
| 196390 | 101905 | $2.12^{+0.01}_{-0.01}$ | $2.15^{+0.01}_{-0.01}$ | $0.088 \pm 0.004$ | 0.019 | $1.07^{+0.02}_{-0.01}$ | $1.40^{+0.30}_{-0.25}$ | $5906 \pm 5$ | $4.50 \pm 0.01$ | $1.23 \pm 0.02$ |
| 197027 | 102152 | $0.58^{+0.01}_{-0.22}$ | $0.63^{+0.01}_{-0.22}$ | $-0.016 \pm 0.003$ | 0.028 | $0.97^{+0.01}_{-0.01}$ | $8.40^{+0.23}_{-0.24}$ | $5718 \pm 4$ | $4.33 \pm 0.01$ | $1.15 \pm 0.02$ |
| 197076 | 102040 | $2.16^{+0.01}_{-0.01}$ | $2.17^{+0.01}_{-0.01}$ | $-0.080 \pm 0.003$ | 0.018 | $1.01^{+0.01}_{-0.01}$ | $2.60^{+0.32}_{-0.27}$ | $5853 \pm 4$ | $4.48 \pm 0.01$ | $0.99 \pm 0.01$ |
| 200633 | 104045 | $1.51^{+0.06}_{-0.03}$ | $1.55^{+0.06}_{-0.03}$ | $0.051 \pm 0.003$ | 0.024 | $1.02^{+0.02}_{-0.01}$ | $5.00^{+0.28}_{-0.28}$ | $5826 \pm 3$ | $4.41 \pm 0.01$ | $0.99 \pm 0.01$ |
| 202628 | 105184 | $2.23^{+0.01}_{-0.01}$ | $2.25^{+0.01}_{-0.01}$ | $0.003 \pm 0.004$ | 0.017 | $1.04^{+0.01}_{-0.01}$ | $1.20^{+0.54}_{-0.45}$ | $5843 \pm 6$ | $4.51 \pm 0.01$ | $1.11 \pm 0.01$ |
| 2071 | 1954 | $1.34^{+0.03}_{-0.06}$ | $1.38^{+0.03}_{-0.06}$ | $-0.090 \pm 0.004$ | 0.026 | $0.96^{+0.01}_{-0.01}$ | $4.80^{+0.39}_{-0.35}$ | $5720 \pm 2$ | $4.46 \pm 0.01$ | $1.00 \pm 0.01$ |
| 207700 | 108158 | $0.56^{+0.19}_{-0.25}$ | $0.62^{+0.19}_{-0.25}$ | $0.055 \pm 0.004$ | 0.024 | $1.02^{+0.01}_{-0.01}$ | $8.10^{+0.13}_{-0.24}$ | $5675 \pm 4$ | $4.29 \pm 0.01$ | $1.09 \pm 0.01$ |
| 20782 | 15527 | $0.64^{+0.14}_{-0.12}$ | $0.68^{+0.14}_{-0.12}$ | $-0.064 \pm 0.003$ | 0.024 | $0.98^{+0.01}_{-0.01}$ | $8.10^{+0.23}_{-0.25}$ | $5779 \pm 4$ | $4.33 \pm 0.01$ | $1.04 \pm 0.01$ |
| 208704 | 108468 | $1.10^{+0.06}_{-0.08}$ | $1.13^{+0.06}_{-0.08}$ | $-0.096 \pm 0.004$ | 0.019 | $1.00^{+0.01}_{-0.01}$ | $7.20^{+0.26}_{-0.26}$ | $5841 \pm 4$ | $4.35 \pm 0.01$ | $0.96 \pm 0.01$ |
| 210918 | 109821 | $0.67^{+0.14}_{-0.23}$ | $0.71^{+0.14}_{-0.23}$ | $-0.108 \pm 0.004$ | 0.023 | $0.97^{+0.01}_{-0.01}$ | $8.90^{+0.23}_{-0.25}$ | $5747 \pm 4$ | $4.31 \pm 0.01$ | $1.11 \pm 0.01$ |
| 219057 | 114615 | $1.86^{+0.01}_{-0.01}$ | $1.89^{+0.01}_{-0.01}$ | $-0.063 \pm 0.004$ | 0.019 | $1.01^{+0.01}_{-0.01}$ | $2.90^{+0.33}_{-0.50}$ | $5819 \pm 5$ | $4.51 \pm 0.01$ | $1.03 \pm 0.01$ |
| 220507 | 115577 | $0.16^{+0.00}_{-0.00}$ | $0.21^{+0.00}_{-0.00}$ | $0.013 \pm 0.003$ | 0.021 | $1.02^{+0.01}_{-0.01}$ | $8.70^{+0.22}_{-0.24}$ | $5694 \pm 4$ | $4.26 \pm 0.01$ | $0.98 \pm 0.01$ |
| 222582 | 116906 | $0.74^{+0.01}_{-0.11}$ | $0.78^{+0.01}_{-0.11}$ | $-0.005 \pm 0.003$ | 0.021 | $1.01^{+0.01}_{-0.01}$ | $6.80^{+0.25}_{-0.23}$ | $5790 \pm 3$ | $4.37 \pm 0.01$ | $1.06 \pm 0.01$ |
| 223238 | 117367 | $1.42^{+0.03}_{-0.08}$ | $1.45^{+0.03}_{-0.08}$ | $0.024 \pm 0.003$ | 0.019 | $1.04^{+0.01}_{-0.01}$ | $5.90^{+0.26}_{-0.23}$ | $5867 \pm 3$ | $4.35 \pm 0.01$ | $0.99 \pm 0.01$ |
| 224383 | 118115 | $0.92^{+0.04}_{-0.08}$ | $0.96^{+0.04}_{-0.08}$ | $-0.036 \pm 0.003$ | 0.019 | $1.02^{+0.01}_{-0.01}$ | $8.20^{+0.23}_{-0.25}$ | $5798 \pm 4$ | $4.28 \pm 0.01$ | $1.20 \pm 0.01$ |
| 25874 | 18844 | $0.67^{+0.18}_{-0.27}$ | $0.72^{+0.18}_{-0.27}$ | $0.014 \pm 0.003$ | 0.024 | $1.00^{+0.01}_{-0.01}$ | $7.50^{+0.23}_{-0.25}$ | $5734 \pm 3$ | $4.37 \pm 0.01$ | $1.02 \pm 0.01$ |
| 30495 | 22263 | $2.37^{+0.01}_{-0.01}$ | $2.38^{+0.01}_{-0.01}$ | $0.037 \pm 0.006$ | 0.018 | $1.05^{+0.01}_{-0.01}$ | $0.63^{+0.34}_{-0.34}$ | $5870 \pm 7$ | $4.54 \pm 0.01$ | $0.98 \pm 0.01$ |
| 36152 | 25670 | $1.11^{+0.05}_{-0.06}$ | $1.16^{+0.05}_{-0.06}$ | $0.054 \pm 0.003$ | 0.026 | $1.01^{+0.01}_{-0.01}$ | $5.10^{+0.27}_{-0.54}$ | $5760 \pm 3$ | $4.42 \pm 0.01$ | $1.02 \pm 0.01$ |
| 3821 | 3203 | $2.45^{+0.01}_{-0.01}$ | $2.45^{+0.01}_{-0.01}$ | $-0.050 \pm 0.007$ | 0.015 | $1.04^{+0.01}_{-0.02}$ | $0.53^{+0.34}_{-0.34}$ | $5868 \pm 9$ | $4.54 \pm 0.02$ | $1.20 \pm 0.01$ |
| 39881 | 28066 | $0.71^{+0.05}_{-0.06}$ | $0.74^{+0.05}_{-0.06}$ | $-0.147 \pm 0.003$ | 0.018 | $0.99^{+0.01}_{-0.01}$ | $9.00^{+0.26}_{-0.20}$ | $5742 \pm 4$ | $4.30 \pm 0.01$ | $1.00 \pm 0.01$ |

**Table A2** – *continued*

| HD | HIP | A(Li) LTE (dex) | A(Li) NLTE (dex) | [Fe/H] (dex) | Convective mass (M$_\odot$) | Mass (M$_\odot$) | Age (Gyr) | $T_{\rm eff}$ (K) | log $g$ (dex) | $v_{\rm t}$ (km s$^{-1}$) |
|---|---|---|---|---|---|---|---|---|---|---|
| 42618 | 29432 | $1.21^{+0.04}_{-0.02}$ | $1.25^{+0.04}_{-0.02}$ | $-0.112 \pm 0.003$ | 0.025 | $0.96^{+0.01}_{-0.02}$ | $6.00^{+0.28}_{-0.72}$ | $5762 \pm 3$ | $4.45 \pm 0.01$ | $1.12 \pm 0.01$ |
| 45021 | 30037 | $0.74^{+0.14}_{-0.22}$ | $0.79^{+0.14}_{-0.22}$ | $0.007 \pm 0.003$ | 0.031 | $0.96^{+0.01}_{-0.01}$ | $6.40^{+0.26}_{-0.42}$ | $5666 \pm 3$ | $4.42 \pm 0.01$ | $1.10 \pm 0.01$ |
| 45289 | 30476 | $0.27^{+0.00}_{-0.00}$ | $0.32^{+0.00}_{-0.00}$ | $-0.033 \pm 0.003$ | 0.023 | $0.99^{+0.02}_{-0.01}$ | $9.40^{+0.19}_{-0.26}$ | $5709 \pm 4$ | $4.28 \pm 0.01$ | $1.00 \pm 0.01$ |
| 45346 | 30502 | $0.95^{+0.10}_{-0.09}$ | $0.99^{+0.10}_{-0.09}$ | $-0.057 \pm 0.004$ | 0.028 | $0.96^{+0.01}_{-0.01}$ | $7.30^{+0.25}_{-0.59}$ | $5731 \pm 4$ | $4.40 \pm 0.01$ | $1.12 \pm 0.01$ |
| 50806 | 33094 | $0.62^{+0.17}_{-0.09}$ | $0.68^{+0.17}_{-0.09}$ | $0.023 \pm 0.005$ | 0.014 | $1.07^{+0.01}_{-0.01}$ | $8.70^{+0.09}_{-0.09}$ | $5629 \pm 7$ | $4.11 \pm 0.02$ | $1.00 \pm 0.01$ |
| 54351 | 34511 | $1.73^{+0.01}_{-0.02}$ | $1.76^{+0.01}_{-0.02}$ | $-0.091 \pm 0.003$ | 0.022 | $0.98^{+0.01}_{-0.01}$ | $5.20^{+0.40}_{-0.46}$ | $5812 \pm 4$ | $4.45 \pm 0.01$ | $1.07 \pm 0.01$ |
| 59711 | 36512 | $1.20^{+0.04}_{-0.04}$ | $1.24^{+0.04}_{-0.04}$ | $-0.126 \pm 0.002$ | 0.026 | $0.95^{+0.01}_{-0.01}$ | $6.30^{+0.29}_{-0.24}$ | $5744 \pm 2$ | $4.45 \pm 0.01$ | $1.05 \pm 0.01$ |
| 59967 | 36515 | $2.68^{+0.01}_{-0.02}$ | $2.67^{+0.01}_{-0.01}$ | $-0.029 \pm 0.009$ | 0.017 | $1.03^{+0.01}_{-0.01}$ | $0.50^{+0.28}_{-0.25}$ | $5855 \pm 12$ | $4.55 \pm 0.02$ | $1.09 \pm 0.01$ |
| 6204 | 4909 | $2.41^{+0.01}_{-0.02}$ | $2.42^{+0.01}_{-0.02}$ | $0.048 \pm 0.006$ | 0.019 | $1.05^{+0.01}_{-0.01}$ | $0.90^{+0.31}_{-0.31}$ | $5861 \pm 7$ | $4.50 \pm 0.02$ | $1.01 \pm 0.01$ |
| 63487 | 38072 | $1.62^{+0.05}_{-0.03}$ | $1.66^{+0.05}_{-0.03}$ | $0.085 \pm 0.007$ | 0.019 | $1.07^{+0.01}_{-0.01}$ | $2.10^{+0.36}_{-0.76}$ | $5860 \pm 9$ | $4.50 \pm 0.02$ | $0.97 \pm 0.01$ |
| 6718 | 5301 | $0.91^{+0.00}_{-0.00}$ | $0.95^{+0.00}_{-0.00}$ | $-0.074 \pm 0.003$ | 0.027 | $0.96^{+0.01}_{-0.01}$ | $6.90^{+0.25}_{-0.24}$ | $5723 \pm 3$ | $4.39 \pm 0.01$ | $0.96 \pm 0.01$ |
| 68168 | 40133 | $1.48^{+0.01}_{-0.02}$ | $1.53^{+0.01}_{-0.02}$ | $0.116 \pm 0.002$ | 0.026 | $1.04^{+0.01}_{-0.01}$ | $5.30^{+0.27}_{-0.24}$ | $5745 \pm 3$ | $4.37 \pm 0.01$ | $0.99 \pm 0.01$ |
| 71334 | 41317 | $0.69^{+0.18}_{-0.19}$ | $0.73^{+0.18}_{-0.19}$ | $-0.081 \pm 0.003$ | 0.028 | $0.95^{+0.01}_{-0.01}$ | $7.70^{+0.32}_{-0.24}$ | $5706 \pm 3$ | $4.38 \pm 0.01$ | $1.02 \pm 0.01$ |
| 73350 | 42333 | $2.25^{+0.01}_{-0.01}$ | $2.28^{+0.01}_{-0.01}$ | $0.132 \pm 0.006$ | 0.022 | $1.07^{+0.01}_{-0.01}$ | $0.90^{+0.29}_{-0.24}$ | $5846 \pm 8$ | $4.50 \pm 0.02$ | $1.09 \pm 0.01$ |
| 75302 | 43297 | $1.59^{+0.01}_{-0.01}$ | $1.64^{+0.01}_{-0.01}$ | $0.082 \pm 0.003$ | 0.028 | $1.01^{+0.01}_{-0.01}$ | $2.60^{+0.26}_{-0.71}$ | $5705 \pm 4$ | $4.50 \pm 0.01$ | $1.02 \pm 0.01$ |
| 78429 | 44713 | $0.59^{+0.14}_{-0.35}$ | $0.64^{+0.14}_{-0.35}$ | $0.063 \pm 0.004$ | 0.023 | $1.03^{+0.01}_{-0.01}$ | $7.60^{+0.27}_{-0.24}$ | $5759 \pm 3$ | $4.28 \pm 0.01$ | $0.99 \pm 0.01$ |
| 78534 | 44935 | $0.98^{+0.06}_{-0.09}$ | $1.02^{+0.06}_{-0.09}$ | $0.038 \pm 0.004$ | 0.024 | $1.01^{+0.01}_{-0.01}$ | $6.60^{+0.26}_{-0.28}$ | $5771 \pm 4$ | $4.37 \pm 0.01$ | $1.23 \pm 0.01$ |
| 78660 | 44997 | $1.14^{+0.04}_{-0.04}$ | $1.18^{+0.04}_{-0.04}$ | $-0.012 \pm 0.003$ | 0.026 | $0.98^{+0.01}_{-0.01}$ | $5.10^{+0.40}_{-0.42}$ | $5728 \pm 3$ | $4.41 \pm 0.01$ | $0.97 \pm 0.01$ |
| 8291 | 6407 | $1.77^{+0.01}_{-0.01}$ | $1.80^{+0.01}_{-0.01}$ | $-0.058 \pm 0.006$ | 0.020 | $1.00^{+0.01}_{-0.01}$ | $2.00^{+0.78}_{-0.57}$ | $5775 \pm 7$ | $4.50 \pm 0.01$ | $0.96 \pm 0.01$ |
| 88072 | 49756 | $1.41^{+0.01}_{-0.03}$ | $1.45^{+0.01}_{-0.03}$ | $0.023 \pm 0.003$ | 0.023 | $1.01^{+0.01}_{-0.01}$ | $5.20^{+0.25}_{-0.37}$ | $5789 \pm 3$ | $4.43 \pm 0.01$ | $1.03 \pm 0.01$ |
| 96116 | 54102 | $2.17^{+0.01}_{-0.01}$ | $2.19^{+0.01}_{-0.01}$ | $0.011 \pm 0.005$ | 0.018 | $1.04^{+0.01}_{-0.01}$ | $1.40^{+0.01}_{-0.58}$ | $5845 \pm 6$ | $4.51 \pm 0.01$ | $1.10 \pm 0.01$ |
| 96423 | 54287 | $1.86^{+0.01}_{-0.01}$ | $1.91^{+0.01}_{-0.01}$ | $0.107 \pm 0.004$ | 0.027 | $1.03^{+0.02}_{-0.01}$ | $5.90^{+0.34}_{-0.25}$ | $5714 \pm 4$ | $4.34 \pm 0.01$ | $1.09 \pm 0.01$ |
| 97037 | 54582 | $1.62^{+0.01}_{-0.02}$ | $1.64^{+0.01}_{-0.01}$ | $-0.096 \pm 0.004$ | 0.016 | $1.02^{+0.01}_{-0.01}$ | $7.70^{+0.26}_{-0.26}$ | $5883 \pm 5$ | $4.28 \pm 0.01$ | $1.04 \pm 0.01$ |
| 9986 | 7585 | $1.79^{+0.01}_{-0.01}$ | $1.83^{+0.01}_{-0.01}$ | $0.083 \pm 0.003$ | 0.023 | $1.04^{+0.01}_{-0.01}$ | $3.70^{+0.28}_{-0.32}$ | $5822 \pm 3$ | $4.45 \pm 0.01$ | $1.02 \pm 0.01$ |

This paper has been typeset from a TEX/LATEX file prepared by the author.